\documentclass[twocolumn]{article}
\usepackage{ijcai23}

\usepackage{times}
\usepackage{soul}
\usepackage{url}
\usepackage[hidelinks]{hyperref}
\usepackage[utf8]{inputenc}
\usepackage[small]{caption}
\usepackage{graphicx}
\usepackage{amsmath}
\usepackage{amsthm}
\usepackage{booktabs}
\usepackage{algorithm}
\usepackage{algorithmic}
\usepackage[switch]{lineno}

\title{On Autonomous Agents in a Cyber Defence Environment}

\author{
Mitchell Kiely
\and
David Bowman\and
Maxwell Standen \and
Christopher Moir
\affiliations
Defence Science \& Technology Group
\emails
\{mitchell.kiely, david.bowman, maxwell.standen1, christopher.moir\}@defence.gov.au
}

\begin{document}

\maketitle

\begin{abstract}
  Autonomous Cyber Defence is required to respond to high-tempo cyber-attacks. To facilitate the research in this challenging area, we explore the utility of the autonomous cyber operation environments presented as part of the Cyber Autonomy Gym for Experimentation (CAGE) Challenges, with a specific focus on CAGE Challenge 2. CAGE Challenge 2 required a defensive Blue agent to defend a network from an attacking Red agent. We provide a detailed description of the this challenge and describe the approaches taken by challenge participants. From the submitted agents, we identify four classes of algorithms, namely, Single-Agent Deep Reinforcement Learning (DRL), Hierarchical DRL, Ensembles, and Non-DRL approaches. Of these classes, we found that the hierarchical DRL approach was the most capable of learning an effective cyber defensive strategy. Our analysis of the agent policies identified that different algorithms within the same class produced diverse strategies and that the strategy used by the defensive Blue agent varied depending on the strategy used by the offensive Red agent. We conclude that DRL algorithms are a suitable candidate for autonomous cyber defence applications.
\end{abstract}

\section{Introduction}
Malicious cyber attacks are becoming increasingly autonomous as our society steers towards a progressively data-driven world. The increasing employment of Internet of Things (IoT) devices across a vast range of sectors, such as healthcare, finance, government has resulted in an increase in potential cyber vulnerabilities. Machine Learning (ML) techniques have been utilised in an attempt to minimise the damage posed by these vulnerabilities. Numerous supervised and unsupervised techniques have been utilised to develop defensive cyber applications such as classifying a type of attack \cite{ch2020computational} and intrusion detection systems \cite{almseidin2017evaluation} \cite{vinayakumar2019deep}. A drawback to the application of these techniques is that they are unable to provide an effective defensive countermeasure to the identified attacker.  Reinforcement Learning (RL), specifically Deep Reinforcement Learning (DRL), has the potential to learn how to identify if an attacker is on a system, as well as actively respond in a way that minimises the damage caused by an attacker. DRL is one approach that may assist in tackling the challenge of evolving autonomous cyber attacks.

Autonomous Cyber Operations (ACO) involves the development of a defensive Blue team and an offensive Red team in an adversarial scenario. In the context of DRL, this may involve training a Blue defensive agent in an environment where it must defend a cyber security system. DRL has the potential to assist in solving many complex ACO problems because of its innate ability to deal with high-dimensional state spaces that are prevalent in many security systems. DRL requires environments in which the algorithms can be trained and tested. \cite{nguyen2021deep} identified that the application of DRL to the cyber security domain falls into two categories: 1) enhancing network capabilities in IoT applications, and 2) defending  against cyber attacks. For the second category, understanding which DRL techniques produce effective defensive protocols and why, is a significant challenge that requires the comparison of a range of algorithms over numerous environments. The open-source Cyber Autonomy Gym for Experimentation (CAGE) \cite{cage_challenge_2} Challenges provides a series of scenarios in which different DRL algorithms may be tried and tested.

The CAGE Challenges encourage participants to compete to produce the most effective algorithm for a given scenario and challenge problem. These challenges provide 2 different classes of ACO problems to solve: single-agent, and multi-agent. Single-agent problems involves  one autonomous decision making agent that selects a single action which is to be carried out at one time step. On the other hand, the multi-agent problem involves numerous actors that must individually select their own actions which are carried out in parallel, or sequentially, to actions selected by other agents. In the multi-agent case, the action selected by one agent may effect the action selected by another.

Other cyber security gym environments, such as \cite{andrew2022developing}, \cite{molina2021network}, and \cite{msft:cyberbattlesim}, offer insight into particular areas of autonomous cyber operations. However, there exists no openly-available systemic classification and comparison of the different DRL and non-DRL classes of algorithms in these environments, or to the challenging problem of ACO in general. Further, an analysis of the strengths and weaknesses of each algorithmic class is entirely absent when taken into the context of cyber security. We envisage the CAGE Challenges to fill in this gap by evaluating each class of algorithms using the same evaluation protocol. Additionally, we highlight situations where certain scenarios are beneficial for particular algorithmic classes as well as identifying where they fail. Another aim of these challenges is to foster a community of like minded individuals that are attempting to solve the difficult problem of autonomous cyber defence as well as enhancing the awareness of DRL applications to the field of cyber security.

In this paper, we present a set of environments for the application of DRL algorithms, known as the CAGE challenges. We define a taxonomy of solution AI algorithms specifically to CAGE challenge 2. This taxonomy permits an inter-group as well as an intra-group comparison between all the algorithmic classes, allowing us to understand and analyse common strengths and weaknesses of the different approaches.

This paper is structured as follows. Section 2 details other publicly available cyber security challenges. Section 3 describes the CAGE Challenge 2 environment. Section 4 proposes an agent categorization system that clusters agent types into classes. Section 5 presents the results of the agent submissions, and analyses the policy that each agent learnt and implemented. Section 6 highlights the strengths and weaknesses of each class of algorithm and provides insight into the importance of how an algorithm is trained within a cyber security scenario.

\section{Related Work}
There is a plethora of public machine-learning based challenges that are designed to foster a community which is aimed at furthering a specific part of a research domain or application. The recent development of DRL has sparked a number of reinforcement learning challenges, such as the Power Network Challenge \cite{kelly2020reinforcement}, the MineRL Competitions \cite{guss2019minerl}, the Learn to Walk Challenge \cite{haarnoja2018learning}, and many more. One objective of these challenges is to catalyse the application of DRL to their respective field. Such a publicly available competition does not exist in the field of cyber security. 

Most publicly available security problems are centered around data-privacy, such as the Hide-and-Seek privacy challenge \cite{jordon2021hide}, or detection/classification problems, such as the Industial Internet of Things (IIoT) Cyber Security HAICK 2022 Challenge \cite{haiick}. The Hide-and-Seek challenge is focused on the healthcare sector which aims to further techniques that are able to generate high fidelity data that minimizes the privacy risk of patient re-identification. On the other hand, The Trojan Detection challenge aims to develop methods which are capable of detecting Trojan attacks on neural networks. This is similar, albeit slightly different to the IIoT HAICK challenge where the aim is to detect and classify the type of attack that has been carried out on an IoT device. In these security challenges, it is more intuitive to use ML techniques that are not RL based. This is the key distinction from the CAGE Challenges, where we aim to inspire the application of  reinforcement learning algorithms, to the complex problem of defensive cyber operations. 

\section{The CAGE Challenges}
The CAGE challenges simulate a range of cyber security scenarios that aim to facilitate research into developing autonomous cyber operators via the application of reinforcement learning. Each challenge is a unique cyber security problem with Red agents crafted for a specific scenario. The purpose of the challenges is to build a Blue agent that is capable of effectively defending the network against the Red agents. A key component of the CAGE challenges is the open-source Cyber Operations Research Gym (CybORG) tool \cite{standen2021cyborg} which can train both single agent and multi-agent reinforcement learning algorithms. The focus of the CAGE challenges is the development of defensive Blue agents that are capable of learning and implementing an effective strategy that counters numerous offensive Red agents. There currently exists three public challenges; the first and second involved training a defender that is capable of defending a small enterprise network, and the third involves training multiple defenders to protect an ad-hoc drone network. This paper will focus on CAGE challenge 2 as it builds on the first challenge and is more complex, and the third challenge only recently concluded.

\subsection{CAGE Challenge 2}
\begin{figure}[H]
    \centering
    \includegraphics[scale=0.28]{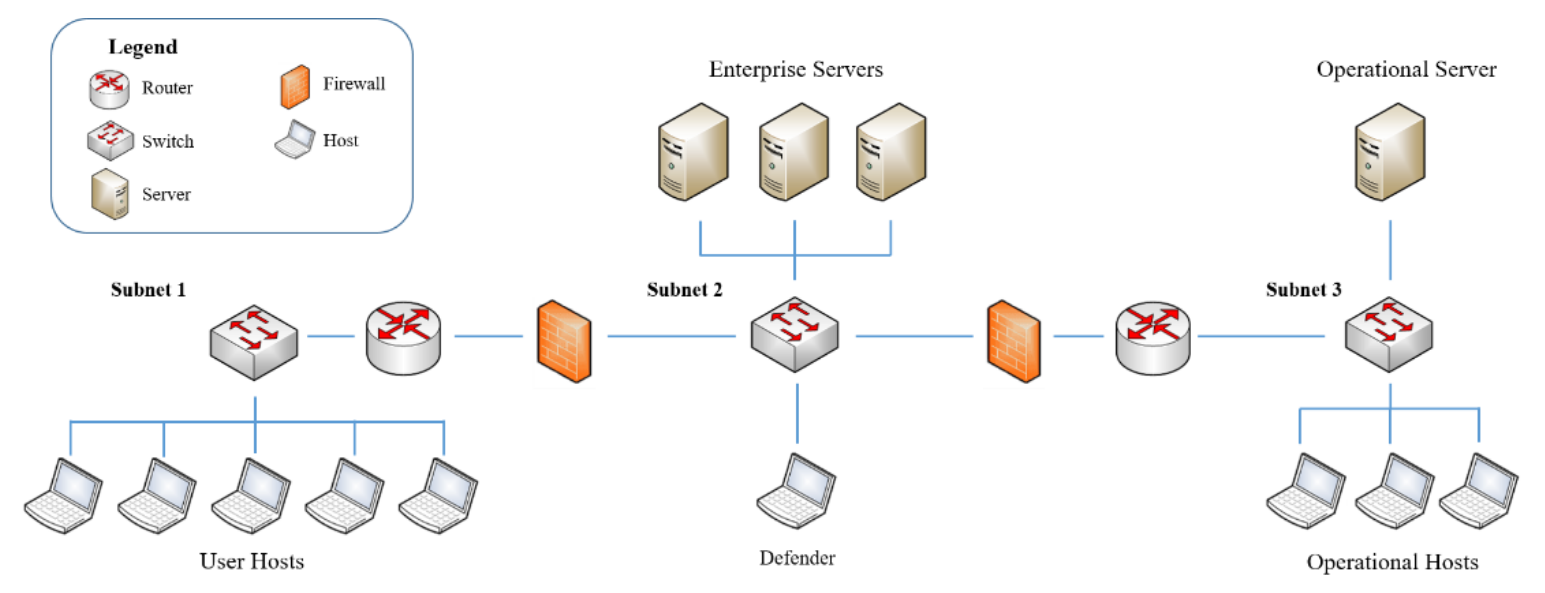}
    \caption{The enterprise network structure of CAGE Challenge 2} \cite{cage_challenge_2}
    \label{fig:cc2}
\end{figure}

CAGE challenge 2 \cite{cage_challenge_2} focused on developing a defensive Blue agent that was able to defend a small enterprise network. The purpose of the defender is to support operations within a factory, the network being defended is displayed in Figure \ref{fig:cc2}.

This challenge extends the previous challenge by greatly increasing the size of the action space, making it more difficult for the Blue agent to discover the optimal strategy. The network is comprised of three subnets; Subnet 1 contains non-critical hosts, Subnet 2 contains servers that support the activities of Subnet 1 as well as a host that represents where a defender may carry out their actions, and Subnet 3 contains the three operational hosts and an operational server that must be maintained in ordered for the network to function properly.

This challenge simulates a post-exploitation lateral movement scenario where the Red agent commences each episode with root access on one user host in Subnet 1. The objective of the Red agent is to reach the operational server and begin to degrade network services. Accordingly, the Blue agent's objective is to minimise the presence of the Red agent whilst maintaining network functionality. 

CAGE Challenge 2 is a partially observable stochastic environment that is described by the tuple $\langle I, {S}, {A}, {O}, T, R_i \rangle$. In this framework, $I = \{1,...,n\}$ is the set of all agents indexed. ${S}$ is a finite set of states, which contains $s_0$, the initial state. ${A} = \{A_1,...,A_n\}$ is a set of the joint action space of each agent $i$ where $A_{i} = \{a_{ix},...,a_{iy}\}$ is the set of actions that an agent $i$ can select. The joint action, $\vec{a} = \langle a_{1x}, ... , a_{ny} \rangle$ is a vector containing the actions chosen by each agent for a given time step. ${O_i}$ is a finite set of observations for each agent $i$. The joint observation space, $\vec{o} = \langle o_1, ..., o_n \rangle$ contains the observation given to each agent. $T$ is the transition probabilities between states where $T(s', \vec{o} | s, \vec{a}$) denotes the probability to transition to state $s'$ producing the joint observation $\vec{o}$ given the current state $s$ and the joint action $\vec{a}$. $R_i$ is the reward function for agent $i$.

There are three unique rules-based Red agents in CAGE Challenge 2: B-line, Meander, and Sleep.  The B-line agent acts with prior knowledge of the network, and selects actions that result in impacting the operational server in the shortest number of steps. The Meander agent has no knowledge of the network layout, and methodically explores and attempts to gain privileged access on all the hosts within each subnet before progressing onto the next subnet. The Sleep agent is a baseline agent that takes no offensive actions. The purpose of the Sleep agent is to test to see if the Blue defender is capable of identifying when there is no threat on the system within the partially observable framework.

\subsection{Action Space and Agent Interaction}
\begin{figure}[H]
    \centering
    \includegraphics[scale=0.35]{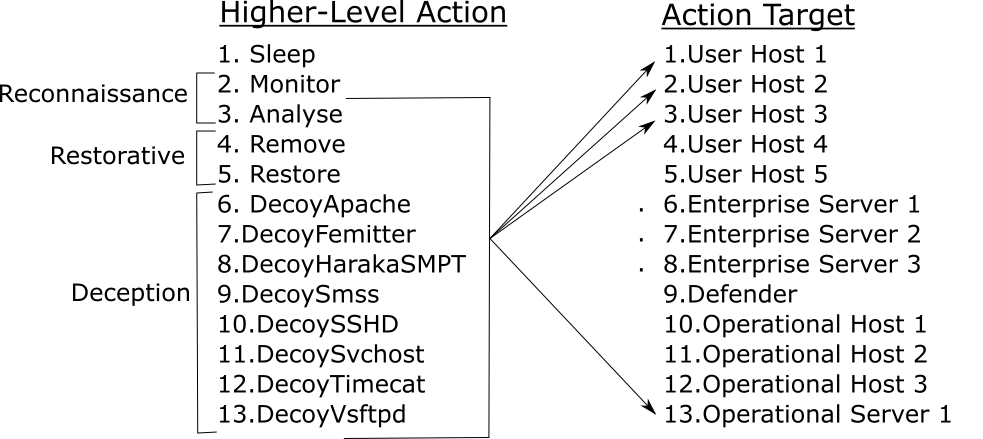}
    \caption{A list of all the higher-level actions. \emph{Sleep} and \emph{Monitor} require no target specification. Actions 3-13 requires an action target, i.e. a host/server to act upon.}
    \label{fig:action_space}
\end{figure}

The action space for the Blue agent to be trained is a discrete set of 145 different actions. At each time step, only one of these actions can be selected by the Blue agent. These actions are outlined in Figure \ref{fig:action_space}. The Red agent's action space is variable with time, as the actions available to the agent vary depending on the amount of access Red has on the network. 

The only common action that both agents can do is the \emph{Sleep} action. If this action is selected by both agents, then there will be no change to the true state of the environment. The action space of the Blue agent (excluding the \emph{Sleep} action) within each environment  can be grouped into three distinct classes: reconnaissance, deception, and restorative. Reconnaissance actions, \emph{Monitor} and \emph{Analyse}, assist the Blue agent in gaining a greater understanding of the current state of the network. \emph{Monitor} provides the Blue agent with a broad sweep over the entire network, whereas \emph{Analyse} gives an in-depth view of one particular host/server. Deceptive actions, these being \emph{DecoyApache}, \emph{DecoyFermitter}, \emph{DecoyHarakaSMPT}, \emph{DecoySmss}, \emph{DecoySSHD}, \emph{DecoySvchost}, \emph{DecoyTomcat} and \emph{DecoyVsfvtp}, lay traps that slow down the movement of the Red agent. Different hosts are configured using different operating systems and have different ports available. Accordingly some \emph{Decoy} actions are ineffective on certain hosts/servers. Restorative actions, \emph{Remove} and \emph{Restore}, return the host or server back to a state where there is no presence of the Red agent. The \emph{Remove} action will only be successful if the Red agent has exploited the host/server, but has not gained root access, otherwise the the action will fail. This differs to the \emph{Restore} action which will successfully remove the Red agent even if it has root access. 

Likewise, the actions of the Red agent can be classified into three categories: Reconnaissance, Exploitation, and Effect. Reconnaissance actions, \emph{DiscoverRemoteSystems} and \emph{DiscoverNetworkServices}, assist the Red agent in learning the network layout and features. The \emph{DiscoverRemoteSystems} action gives the Red agent knowledge about the undiscovered subnet, this allows the Red agent to observe the current hosts/servers that are on a subnet. The \emph{DiscoverNetworkServices} gives the Red agent knowledge about a  host/server, this could include the Operating System, the ports which are open, and much more. Exploitation actions, \emph{ExploitRemoveServer} and \emph{PrivilegeEscalate}, utilise the knowledge obtained from the reconnaissance actions to expand the Red agent's presence on the network. \emph{ExploitRemoveServer} exploits a vulnerability identified in the host/server which was uncovered by the \emph{DiscoverNetworkServices} action. This action has a small chance of giving the Red agent root access on the host/server. After successfully exploiting a vulnerability on host/server, the Red agent is then capable of selecting the\emph{PrivilegeEscalate} action. This action utilises this exploited vulnerability to obtain root access on the host/server without any chance of failing. This ensures that the agent is guaranteed to obtain root access unless the Blue agent removes the Red agent before this action is selected.  The purpose of the Effect action, \emph{Impact}, is to disrupt the network's functionality. This action represents the worse case scenario for the Blue agent where the Red agent is actively disrupting the operations being carried out on the network.

\begin{figure}[H]
    \centering
    \includegraphics[scale=0.35]{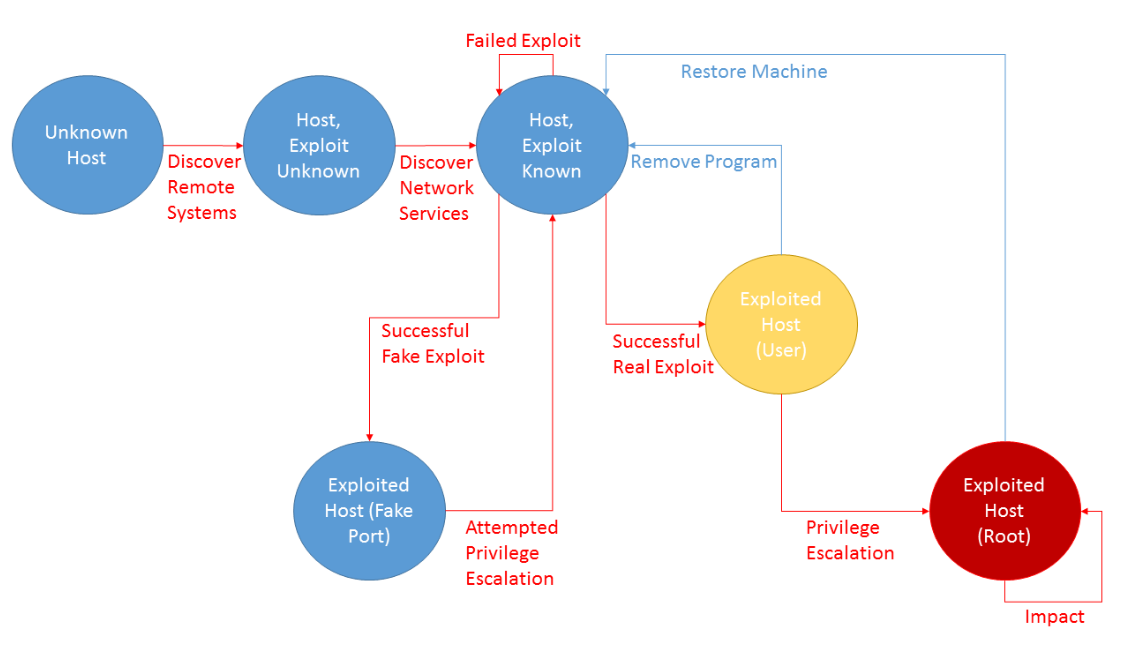}
    \caption{State-transition diagram for CAGE Challenge 2. The circles indicate all the potential states of a host/server. The red arrows show the kill chain of the Red agent. The defensive actions that the Blue agent can select are indicated by blue arrows.}
    \label{fig:state_trans}
\end{figure}

A successful Red kill chain involves five steps: 1) Discovering the users/hosts on a particular subnet; 2) Discovering the services  for a given host/server within the subnet; 3) Exploiting a vulnerability that was discovered on a host/server; 4) Escalating the Red agent's privilege to obtain root access on a host/server; and 5) Impacting the host/server.  The state-transition diagram depicting the kill chain of the Red agent, and the preventative actions that the Blue agent can select, is shown in Figure \ref{fig:state_trans}.
 
\subsection{Reward Function and Evaluation}
 The reward structure reflects the Blue agent's objective goal by 1) Penalising the Blue agent for each host/server that the Red agent has root access on, 2) Penalising the Blue agent for undertaking the costly \textit{Restore} action, and 3) Penalising the Blue agent whenever the Red agent impacts the operational server. Mathematically, this is given as:
 
\begin{align*}
    R_{Blue} &= -0.1(host_{Red} - 1) - server_{Red}  \\
    &  - Blue_{Restore} - 10(Red_{impact})
    \label{reward}
\end{align*}

where $R_{Blue}$ is the reward for the Blue agent, and $host_{Red}$ and $server_{Red}$ are the number of hosts and servers that the Red agent has root access on respectively. The $-1$ term is a result of the Red agent having a foothold on the User 0 host at the commencement of each episode, and the Blue agent is unable to remove the Red agent from this host. $Blue_{restore}$ is 1 if the Blue agent has executed the restore action and 0 otherwise. $Red_{impact}$ is 1 if the Red agent successfully executed the impact action on the operational server and 0 otherwise. The reward for the Blue agent is calculated at every time step.

The theoretical maximum reward value that the blue agent can obtain is zero, where the Red agent does not gain access to any other user or server, and the Blue agent never selects the \textit{Restore} action. An agent that consistently obtains a higher reward than other agents has learnt and implemented a more effective policy. Each submitted Blue agent was evaluated against each rules-based Red agent, for a variable number of time steps of 30, 50, and 100. Each pairing of Red agent and time step was simulated for a total of 1000 episodes. Each episode terminated after the specified number of time steps had been reached, and the cumulative reward, which is the sum of every reward calculation taken in a given episode, is outputted. The actions chosen by the Blue and Red agents at every time step are recorded as part of the evaluation process.

Using the catalogue of Blue agents submitted by challenge participants, we constructed a taxonomy of algorithmic classes that enabled us to more readily compare each Blue agent within its own classification, as well as across classifications. This ultimately assists future attempts to create an agent that will be most capable of defending a network against.

\section{Agent Categorisation}
A total of eighteen agents that were submitted to CAGE Challenge 2, of which sixteen were able to be evaluated. The submitted agents were categorised based on the ML approach utilised, resulting in an agent being assigned into one of the following broadly defined agent classes: 1) Single-Agent DRL; 2) Hierarchical DRL; 3) Ensembles; and 4) Non-DRL approaches. Categorising  into one of the aforementioned classes permitted a richer analysis of agents both within a class, as well as across classes. The taxonomy of agents, and where they fit relative to each other, is shown below in Figure \ref{fig:agent_tax}.

\begin{figure}[H]
    \centering
    \includegraphics[scale=0.3]{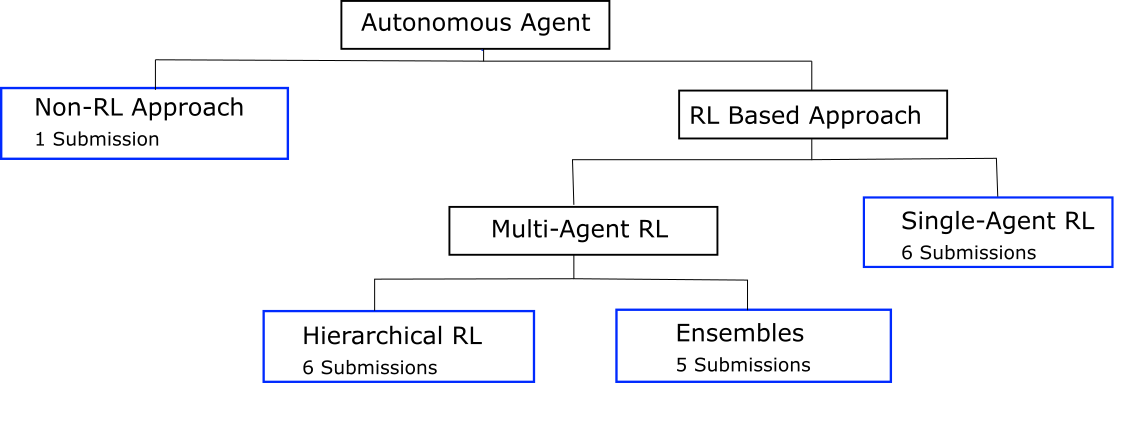}
    \caption{A taxonomy of agent classes and the number of agent submissions for each class that we analysed for CAGE Challenge Two.}
    \label{fig:agent_tax}
\end{figure}

\subsection{Single-Agent DRL}
This class contains the simplest architecture and can be easily implemented through utilising a standard off-the-shelf algorithm, such as a Proximal Policy Optimastion (PPO) \cite{schulman2017proximal}. Single-Agent DRL approaches involve agents that are composed of a single policy network that learns and implements a policy in the given environment. Further, the same neural network is used to evaluate the submission across all nine trials. 

\subsection{Hierarchical DRL}
Hierarchical DRL algorithms are an agent with multiple policy networks and a \textit{selector agent} to choose between them. An example of this approach is to use one selector agent, and three PPO algorithms that were trained against different Red agents. This is a more targeted approach as the task of defending the network is split up into a sequence of sub-tasks where each agent within the hierarchy is responsible for completing each sub-task.

The job of the selector agent is to identify which Red agent is currently attacking the system. Once the attacker has been identified the most appropriate defender  is selected, i.e. the Blue agent that has been specifically trained against the identified Red agent. At each time step, the selected defender determines what action is taken.

\begin{table*}[t]
\caption{95\% confidence intervals for each submitted agent are present at each of the evaluation's duration. The terminology used is as follows: \textit{HPPO} is a Hierarchical Proximal Policy Optimisation approach, \textit{Ens} is an ensemble methodology, \textit{Heur} is the inclusion of a heuristic or rules-based agent, \textit{TL} utilised transfer learning, \textit{Masking} involved using an action mask.}
\label{sample-table}
\resizebox{\textwidth}{!}{%
  
  \begin{tabular}{lllllll}
    \toprule
    \multicolumn{7}{c}{\hspace{45 mm}Episode Length \& Red Agent}                   \\
    \cmidrule(r){2-7}
    \multicolumn{7}{c}{ \hspace{40 mm}30 timesteps  \hspace{30 mm}  50 timesteps   \hspace{35 mm}  100 timesteps}      
    \\
    \cmidrule(r){2-3}
    \cmidrule(r){4-5}
    \cmidrule(r){6-7}
    \\
        
    Blue agent     &  B-Line    &  Meander  &  B-Line    &  Meander & B-Line    & Meander \\
    \midrule
    Agent 1: HPPO + Heur & [-3.57, -3.36]  &  \textbf{[-5.72, -5.56]} & [-6.56, -6.25] & \textbf{[-8.82, -8.56]} & [-14.01, -13.51] & \textbf{[-16.84, -16.35]}     \\
    Agent 2: Ens of Ens PPO & [-3.59, -3.36]  &  [-6.57, -6.37] & [-6.05, -5.65] & [-10.51, -10.16] & [-11.78, -11.01] & [-19.71, -19.05]         \\
    Agent 3: HPPO & [-3.64, -3.48]  &  [-6.84, -6.67] & [-6.41, -6.00] & [-10.27, -9.99] & [-13.17, -12.39] & [-17.94, -17.36]        \\
    Agent 4: HPPO + Heur   & [-3.82, -3.55]  &  [-6.90, -6.72] & [-6.47, -6.02] & [-10.23, -9.94] & [-l3.26, -12.46] & [-17.89, -17.31]            \\
    Agent 5: HPPO  & [-3.97, -3.47]  &  [-6.94, -6.76    ] & [-7.14, -6.17] & [-10.11, -9.84] & [-l3.26, -12.46] & [-17.63, -17.14]            \\
    Agent 6: Ens PPO & [-3.72, -3.50]  &  [-6.67, -6.46] & [-6.30, -5.91] & [-10.50, -10.15] & [-12.28, -11.53] & [-20.53, -19.77]               \\
    Agent 7: Belief Ens PPO & [-3.99, -3.75]  &  [-6.66, -6.42] & [-6.90, -6.57] & [-12.28, -11.74] & [-14.95, -14.30] & [-26.83, -23.35]    \\
    Agent 8: PPO TL & [-5.71, -5.36]  &  [-6.51, -6.29] & [-10.03, -9.45] & [-10.37, -10.04] & [-20.00, -19.14] & [-19.05, -18.26]                \\
    Agent 9: PPO + Masking & [-4.82, -4.38]  &  [-6.55, -6.35] & [-8.32, -7.51] & [-11.46, -10.99] & [-18.95, -16.54] & [-25.14, -23.30]               \\
    Agent 10:PPO  & [-4.93, -4.58]  &  [-8.17   ,-7.90] & [-8.42, -7.83] & [-13.19, -12.63] & [-17.84, -16.55] & [-23.45, -22.44]             \\
    Agent 11: HPPO & [-4.54, -4.34]  &  [-6.71, -6.43] & [-7.83, -7.56] & [-12.10, -11.45] & [-15.90, -15.47] & [-30.74, -28.31]              \\
    Agent 12: Ens PPO & [-6.20, -5.84]  &  [-6.34, -6.12] & [-10.47, -9.92] & [-11.98, -11.5] & [-22.63, -21.54] & [-26.05, -25.04]             \\
    Agent 13: PPO  & [-9.03, -8.47]  &  [-7.41, -7.21] & [-15.37, -14.46]& [-12.30, -11.78] & [-31.26, -29.12] & [-22.79, -21.63]            \\
    Agent 14: HPPO & [-4.74, -4.52]  &  [-9.26, -9.04] & [-8.69, -8.21] & [-22.65, -21.95] & [-17.75, -17.12] & [-59.25, -56.19]                \\
    Agent 15: ACME IMPALA & [-6.25, -5.74]  &  [-9.36, -9.17] & [-10.69, -9.81] & [-22.48, -22.06] & [-20.53, -18.90] & [-55.28, -54.29]             \\
    Agent 16: PPO + RE3 & [-13.29, -11.68]  &  [-7.11, -6.87] & [-24.76, -21.89] & [-12.94, -11.58] & [-56.88, -51.14] & [-30.62, -26.75]              \\
    Agent 17: Ens DDDQN & [-6.12, -5.61]  &  [-11.10, -10.70] & [-11.26, -10.37] & [-26.76, -25.82] & [-25.08, -22.84] & [-69.31, -66.47]             \\
 
    \bottomrule
  \end{tabular}}
  \label{results_table}
\end{table*}

\subsection{Ensembles}
An ensemble approach coalesces numerous individually trained algorithms into a single agent \cite{wiering2008ensemble}, i.e., a set of policy networks. After receiving an observation, the agent passes it into every policy network and receives a set of corresponding action choices. The most frequently selected action is then returned to the environment. Individually trained policy networks within the ensemble can be derived from either algorithms trained against the same Red agent, or the same algorithm trained against a different Red agent.

\subsubsection{Ensemble of Ensembles}
In an ensemble of ensembles approach, the agent consists of a set of ensembles. Each ensemble  makes one action choice and the most frequently selected action across all the ensembles is returned to the environment.  

\subsection{Non-DRL Based Approaches}
This category encapsulates all other agent approaches which do not utilise any form of deep neural network. Just one submission was received into this category, which utilised a Monte-Carlo Tree Search \cite{chaslot2008monte} (MCTS) algorithm. This approach scored the lowest, and was also unable to go through the entire evaluation process due to time constraints. Only two episodes, for each agent-duration pair were able to be carried out within a reasonable time frame. As such, the results have been excluded from this paper but can be found on the CAGE Challenge 2 GitHub page\cite{cage_challenge_2}.

\section{Results \& Discussion }
In general, agents from the Hierarchical DRL class performed best on most evaluation trials. The next best class of agents utilised Ensemble approaches, followed by the Single-Agent DRL class. Additionally, there was a great diversity of rewards obtained within each class. For example, Agent 1 and Agent 14 are both Hierarchical DRL approaches, however the difference in the confidence interval against the Meander agent for 100 time steps is relatively large.

No submitted agent obtained the highest reward across every agent-duration trial. Agent 1 was the most effective at defending the network against the Red Meander agent. However, the defensive policy learnt by Agent 1 against the B-Line agent was inferior compared to the strategy implemented by Agent 2. 
 
We found that almost all the agents obtained the theoretical highest reward, i.e. a reward of $0$ with no variance, against the Sleep agent. This means that the DRL agents were able to recognise that the Sleep agent posed no threat, as the Blue agent did not select the Restore action on any user/host. For brevity, we chose not to display these confidence intervals in the table but they can be found on the CAGE Challenge 2 GitHub page.

\begin{figure*}[t]
    \centering
    \includegraphics[scale=0.45]{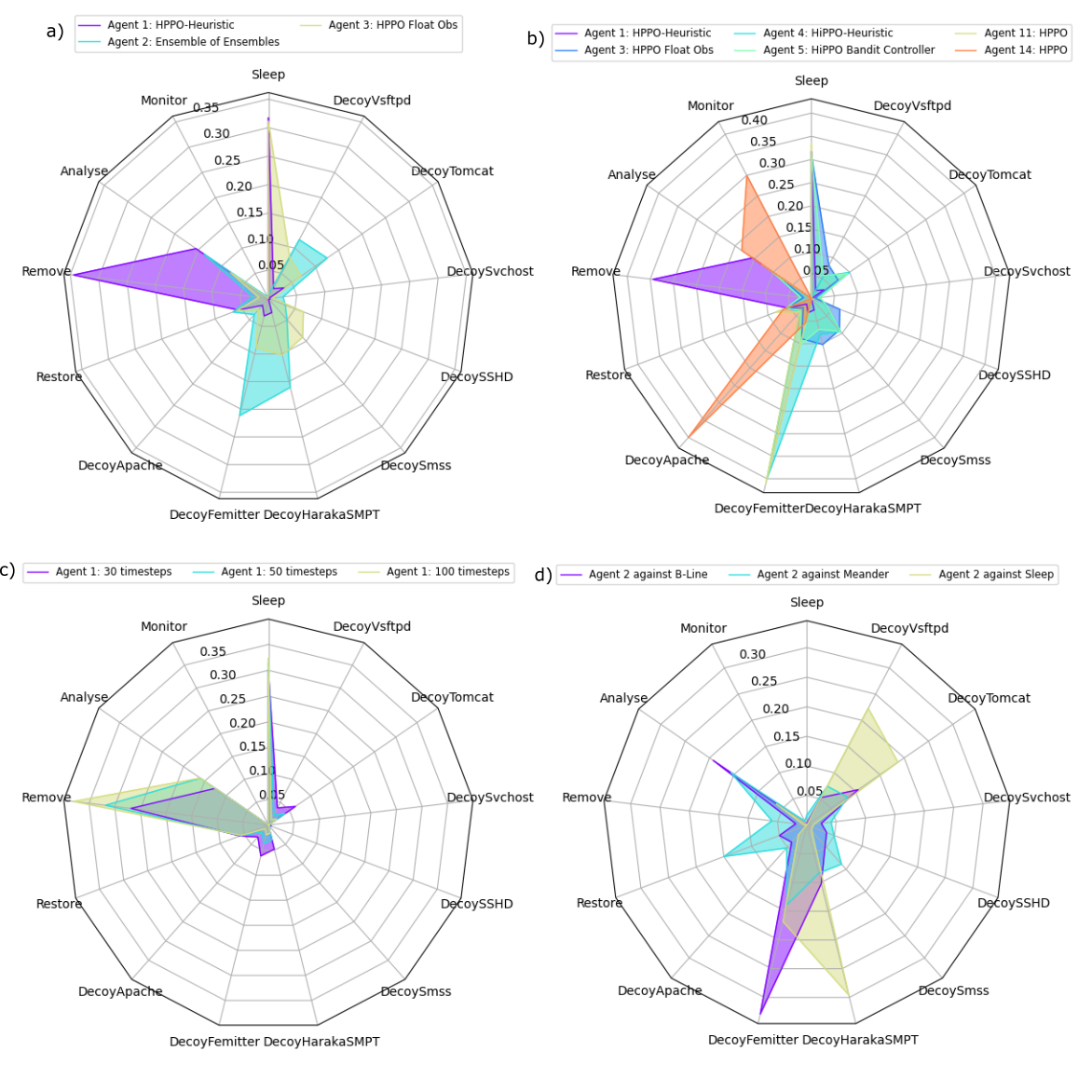}
    \caption{The action-type selection percentage for a variety of agents across a range of scenarios. a) Strategies implemented by Agents 1-3 averaged over evaluation scenarios. b) The strategies learnt by each agent within the hierarchical class. c) The strategy of Agent 1 with different episode time step lengths. d) The strategy Agent 2 learnt for each Red agent.}
    \label{fig:action_sel}
\end{figure*}
The PPO algorithm was the most commonly used algorithm that was submitted to CAGE Challenge 2. Agents 1-6 exclusively utilised PPO, four which came from the Hierarchical DRL class, which we labelled as HPPO. Five out of the six single-agent DRL agents (Agents 8,9,10,13,16) implemented a PPO algorithm but each scored remarkably different results. Agents which did not utilise the PPO algorithm, such as Agent 15 that implemented an ACME Impala \cite{hoffman2020acme}, and Agent 17 that used DDDQN \cite{wang2016dueling}, typically scored lower than agents that utilised the PPO algorithm.

\subsection{Agent Policy Analysis}
To visualise and compare across the various agent policies we developed radar charts presented in Figure 5, which display the distribution of the higher level actions selected by each agent. These radar charts detail the percentage of action types selected by an agent, thus providing us with a high-level view of the agents’ policy for a given scenario. These charts permit a comparative policy analysis between different Blue agents as well as the comparison of a single Blue agent’s strategy in different scenarios such as facing different Red agents. While these graphs do enable a high level comparison of policies, they are unable to capture specific behaviours of an agent. For example, Agent 1 and Agent 5 appear to have very similar strategies, however without state or observation information we are unable to conclude that the strategies are actually similar.

Agents 1-3 learnt vastly different strategies that were all highly capable of defending the enterprise network. This is shown in Figure \ref{fig:action_sel}(a), where Agent 2 and Agent 3 employed a strategy that deployed many decoys on the network, which differs to Agent 1's analyse and remove strategy. Although Agent 2 (Ensemble) and Agent 3 (Hierarchical DRL) both appear to utilise decoys, they differ in the exact ratio of decoys used. While Agent 3 used a wider variety of decoys which encompasses six of the seven decoy actions, Agent 2 focused on using a subset of four of the available seven decoy types. Decoys stall an attacking Red agent resulting in the agent continuously exploiting fake ports within a host. They are an effective countermeasure given it takes the Blue agent only one timestep to set up a Decoy, whereas the Red agent requires a minimum of two timesteps, in most cases, to realise it is dealing with a decoy. However, there is a small chance that the Red Agent gains root access through the \emph{ExploitRemoteServices} action, hence why the average reward returned was not 0. On the other hand, Agent 1 used a lot less Decoys and instead opted to analyse and remove the Red agent before it was able to escalate its privileges. This proved to be a highly effective approach, provided that the Blue agent was able to successfully identify the host/server currently being targeted. All of these effective strategies, however different, achieved a relatively similar reward.

Agents from the same algorithmic class did not necessarily learn the same policy. This can be seen in Figure \ref{fig:action_sel}(b), which presents the action selection percentage for each HPPO algorithm across all the scenarios. Similar to Figure \ref{fig:action_sel}(a), there are a variety of strategies learnt by each agent within the same class. For example, Agent 1 and Agent 14 learnt the most unique strategies with little overlap when compared to other agents from the same class. Agent 1 and 5 have similar distribution of actions, with a focus on selecting the Femitter decoy. This differs from Agent 3, which opts to select a greater variety of decoys to implement. 

The length of the episode impacted the percentage that an action was selected. This is displayed in Figure \ref{fig:action_sel}(c) where the policy snapshot of Agent 1 across the different maximum lengths of episodes is detailed. As can be seen, episodes with a lower step count result in the application of more decoys. As the step count increases, the agent increasingly relies on its ability to identify and remove the Red agent. This suggests that the optimal strategy may change toward the later stages of an episode.

Some agents learnt to implement different strategies against different Red agents. Figure \ref{fig:action_sel}(d) demonstrates that the same agent, in this case Agent 2, has a different action selection percentage when it defends the network against different Red agents. It appears Agent 2 learnt to implement many Femitter decoys against the B-Line agent, but applied a broad sweep of decoys against the Meander agent, which also included a higher number of Restore actions. These strategies align with our understanding of the designed behaviour of the Red agents. For instance, the B-Line agent will persistently attempt its pre-defined strategy. Utilizing deception to counter this strategy results in a Blue agent with a narrow focus on highly successful decoys. The Meander agent uses a more diverse strategy and so it is not unexpected that the blue agent must also use a more diverse strategy.

\subsection{Strengths and Weaknesses of the Agent Classes}
A benefit of the Single-Agent DRL class is that the algorithm can be implemented with little effort. However, a drawback to using this approach is that the Blue agent is forced into one of two outcomes: 1) Training the agent against just one attacker will result in a learnt policy that is only effective in countering that attacker, while performing poorly against other attackers; or 2) Training the agent against all attackers will result in a learnt policy that generalises across all attackers, but the performance against a specific Red agent is inferior to the specialised defender agents described in the aforementioned outcome.

The hierarchical approach does not suffer from the same limitations as the single-agent approach. The purpose of this approach is to specialise different algorithms to learn a policy that specifically counters one particular Red agent.

The Ensemble approach utilises a suite of different method that allows for a broader defence strategy capable of defending a network against a variety of attackers. Implementing this approach requires training multiple defenders, as well as performing an action selection task at each timestep. Thus Ensemble approaches incur a significant increase in computational cost. Furthermore, the time taken to ultimately decide which action to take increases for every individual agent that is a part of the greater ensemble.

\section{Conclusion}
The CAGE challenges are a set of unique cyber challenges that aim to investigate which algorithmic classes are most capable of producing an effective autonomous cyber defensive agent. These challenges allow for an in-depth analysis of the strengths and weaknesses of the agent classes when applied to a cyber security setting. CAGE Challenge 2 is a post-exploitation lateral movement scenario within a small enterprise network that resulted in a myriad of submissions which we classified into four categories: 1) Single-Agent DRL; 2) Hierarchical DRL; 3) Ensembles; and 4) Non-DRL approaches. Of these four, it was clear that the hierarchical DRL approach was the most capable of learning and implementing an effective cyber defensive protocol. We found that the agents used a diverse range of defensive strategies and that different algorithms within the same class often learnt different strategies. Further, the strategy used by a single agent varied depending on the Red agent it was up against. The diversity, and effectiveness of the strategies that the agents produced indicates that DRL algorithms are a suitable candidate for autonomous cyber defence, however, further investigation into more complex scenarios is required. The CAGE challenges aim to foster a community of both ML and Cyber researchers with the goal of developing autonomous agents that are able to select effective defensive actions within a real world network.

\bibliographystyle{named}
\bibliography{acd23}

\end{document}